\pdfoutput=1
\documentclass[10pt,a4paper]{article}

\usepackage[english]{babel}
\usepackage{kantlipsum}
\usepackage[draft]{pgf}
\usepackage{lipsum}
\usepackage{amsmath}
\usepackage{amssymb}
\usepackage{amsthm}
\usepackage{graphicx}
\usepackage[format=hang, font=sf]{caption}
\usepackage[sc]{mathpazo}
\usepackage[T1]{fontenc}
\usepackage{microtype}
\usepackage{breqn}
\usepackage[gen]{eurosym}
\usepackage{cite}
\usepackage{dcolumn}
\usepackage{booktabs}
\usepackage{stfloats}
\usepackage{tikz}
\usepackage[makeroom]{cancel}
\usepackage{cleveref}
\usepackage[numbers]{natbib}
\usepackage{fixltx2e}

\usepackage{todonotes}
\usepackage{url}
\usepackage{algorithmicx}
\usepackage{algorithm}
\usepackage{algpseudocode}

\usepackage{graphicx}
\usepackage{subcaption}
\usepackage{multirow}
\usepackage{xfrac}
\usepackage{mathtools}

\usepackage{array}
\newcolumntype{C}{>{$}c<{$}}

\usetikzlibrary{shapes, arrows}
\tikzset{
    events/.style={ellipse, draw, align=center},
}

\newcommand{\ps}{\textit{Prune Sampling }}
\newcommand{\Ps}{\textit{Prune Sampling }}

\usepackage[hmarginratio=1:1, left=20mm, top=20mm, bottom=33mm, columnsep=20pt]{geometry}
\usepackage{booktabs}
\usepackage{float}

\newcommand{\vs}{\vspace{0.75pc}}

\usepackage{graphicx}
\usepackage{subcaption}
\usepackage{multirow}
\usepackage{xfrac}
\usepackage{mathtools}

\usepackage[explicit]{titlesec}
\usepackage{sgame}
\usepackage[bottom]{footmisc}

\usepackage{soul}

\newcommand\indep{\protect\mathpalette{\protect\independenT}{\perp}}
\def\independenT#1#2{\mathrel{\rlap{$#1#2$}\mkern2mu{#1#2}}}

\newcommand{\bfe}{{\mathbf{e}}}
\newcommand{\bfx}{{\mathbf{x}}}
\newcommand{\bfy}{{\mathbf{y}}}
\newcommand{\bfX}{{\mathbf{X}}}
\newcommand{\bfE}{{\mathbf{E}}}

\newcommand{\C}{{\mathcal C}}
\newcommand{\E}{{\mathbb E}}

\newcommand{\G}{{\mathcal G}}

\newcommand{\I}{{\mathcal{I}}}

\newcommand{\U}{{\mathcal{U}}}
\newcommand{\X}{{\mathcal{X}}}

\usepackage{abstract} 

\renewenvironment{abstract}
 {\small
  \begin{center}
  \normalfont\scshape\bfseries \sffamily \MakeLowercase \abstractname\vspace{-.5em}\vspace{0pt}

  \end{center}
  \list{}{
    \setlength{\leftmargin}{.25cm}%
    \setlength{\rightmargin}{\leftmargin}%
  }%
  \item\relax}
 {\endlist}

\theoremstyle{plain}

\theoremstyle{remark}

\newtheorem*{theorem*}{\textbf{\em Theorem}}

\newtheorem*{remark*}{\textbf{\em Remark}}

\newtheorem{example}{\textbf{\em Example}}
\newtheorem*{example*}{\textbf{\em Example}}

\newtheorem*{lemma*}{\textbf{\em Lemma}}

\newtheorem{definition}{\textbf{\em Definition}}
\newtheorem*{definition*}{\textbf{\em Definition}}

\newtheorem*{proposition*}{\textbf{\em Proposition}}

\newtheorem*{corollary*}{\textbf{\em Corollary}}

\theoremstyle{plain}

\usepackage[T1]{fontenc}
\usepackage[defaultsans,osfigures,scale=0.95]{opensans}
\titleformat{\section}[display]{\normalfont\scshape\bfseries}{}{0em}{\sffamily \MakeLowercase{#1}}
\titlespacing*{\section}{0em}{1.5em}{0em}

\titleformat{\subsection}[display]{\normalfont\scshape}{}{0em}{\sffamily \MakeLowercase{#1}}
\titlespacing*{\subsection}{0em}{1em}{0.2em}

\DeclareCaptionLabelFormat{bsc}{\textbf{\MakeLowercase{\textsc{#1}}\ #2}}

\begin{document}\sloppy

\twocolumn[{
\begin{center}
{\Large{{Prune Sampling}: a MCMC inference technique for discrete and deterministic Bayesian networks}} 
\end{center} \vs

\begin{minipage}[t]{.32\textwidth}
\begin{center}
	{ \textbf{Frank Phillipson} }\\
	{Unit ICT,\\ TNO,\\ the Netherlands} \vs
	
	\textit{frank.phillipson@tno.nl}
\end{center}
\end{minipage}
\hfill
\begin{minipage}[t]{.32\textwidth}
\begin{center}
	{ \textbf{Jurriaan Parie} } \\
	{Mathematical Institute, \\Utrecht University,\\ the Netherlands}\vs
	
	\textit{jfparie@me.com}
\end{center}
\end{minipage}
\hfill
\begin{minipage}[t]{.32\textwidth}
\begin{center}
	{ \textbf{Ron Weikamp} }\\
	{Department of Mathematics,\\ University of Amsterdam,\\ the Netherlands}\vs
	
	\textit{ronweikamp@gmail.com}
\end{center}
\end{minipage} \vs\vs
}] 

\begin{abstract}
We introduce and characterise the performance of the Markov chain Monte Carlo (MCMC) inference method \ps for discrete and deterministic Bayesian networks (BNs). We developed a procedure to obtain the performance of a MCMC sampling method in the limit of infinite simulation time, extrapolated from relatively short simulations. This approach was used to conduct a study to compare the accuracy, rate of convergence and the time consumption of \ps with two conventional MCMC sampling methods: Gibbs- and Metropolis sampling. We show that Markov chains created by \ps always converge to the desired posterior distribution, also for networks where conventional Gibbs sampling fails. Beside this, we demonstrate that pruning outperforms Gibbs sampling, at least for a certain class of BNs. Though, this tempting feature comes at a price. In the first version of \textit{Prune Sampling}, for large BNs the procedure to choose the next iteration step uniformly is rather time intensive. Our conclusion is that \ps is a competitive method for all types of small and medium sized BNs, but -- for now -- standard methods still perform better for all types of large BNs.
\end{abstract}
 
\vspace{-1pc}
\section{Introduction}
Bayesian networks (BNs) are used to model complex uncertain systems with interrelated components. As an illustration, stochastic and deterministic dependencies are used in discrete BNs to model genetic linkage \cite{fishelson2004}, causal reasoning \cite{pearl2014probabilistic} and defence systems \cite{phillipson2015modelling}. The activity of calculating the posterior distribution of a BN given certain evidence is called inference. Exact inference in BNs is often too computationally intensive. On the other hand, popular approximate inference methods often perform poorly in the presence of determinism \cite{koller2009probabilistic, poon2006sound, gogate2011samplesearch}. Due to the real world applications of BNs, improving the reliability of approximate inference methods is quite important and can have a significant impact. Many solutions have been proposed in the past to address this problem. In the section {\normalfont\scshape \sffamily background and notation}, we elaborate further on some of these sampling methods and their pitfalls. 

\indent In this article, our main contribution is the introduction of \textit{Prune Sampling}. This is a Markov chain Monte Carlo (MCMC) sampling method that always converges to the correct posterior distribution, even in the presence of determinism. This technique is inspired by the sound MC-SAT algorithm, which takes advantage of auxiliary variables and slice sampling in the more general framework of Markov Logic Networks (MLNs) \cite{poon2006sound}. Though, \ps avoids the memory intensive translation of a BN to a MLN. Instead, \ps brings a key feature of the MC-SAT algorithm -- the construction of a random sample space --  to the field of BNs. In doing so, it makes use of the compact and graphical structure of BNs directly. 

\indent The key idea of \ps is straight-forward: since the exhaustive listing of all feasible states of the original BN is impossible (due to too much memory and time consumption), the exhaustive listing of all solutions of the randomly pruned BN is possible. The implementation of the pruning technique requires two non-trivial steps: to generate an initial state of the BN and to sample uniformly over the pruned BN. We explain how we met these requirements in this first version of \ps and explain how the MC-SAT algorithm deals with these questions.

\indent We conducted experiments with the class of BNs that the conventional MCMC approximation technique Gibbs sampling fails and explain why \ps does converge to the correct distribution on these BNs. 
Then, we compare the accuracy, the rate of convergence and the time consumption of the \ps algorithm with Gibbs- and Metropolis sampling. For evaluation, we used BNs from several benchmark domains with a gradually increasing level of either available evidence or deterministic relations. Furthermore, we show mathematically why \ps always guarantees convergence of the constructed Markov chain.

\vspace{-1.5pc}
\section{Background and notation}
We start with a brief review of the main concepts we use in this article: the BN framework, the posterior distribution -- which is of interest when doing inference -- and MCMC sampling methods. Beside this, we demonstrate the pitfall of the arguably most widely used MCMC inference technique Gibbs sampling in the presence of determinism and elaborate on the influences the MC-SAT inference algorithm had on the idea of \textit{Prune Sampling}.

\vspace{-1pc}
\subsection{Bayesian networks}

\begin{definition}[Bayesian network]
A BN structure $\G$ is a directed acyclic graph whose nodes represent random variables $\X = (X_1, \ldots , X_n)$. Let $\text{Pa}_{X_i}$ denote the direct parents of $X_i$ in $\G$ and $\text{ND}_{X_i}$ denote the variables in the graph that are non-descendants of $X_i$, where $1 \leq i \leq n$. Then $\G$ encodes the following set of conditional independence assumptions, called the local independencies, and denoted by $\I_l(\G)$:
\begin{align*}
\text{for each variable $X_i$: $(X_i \indep \text{ND}_{X_i} | \text{Pa}_{X_i})$}.
\end{align*}
In other words, the local independencies state that each node $X_i$ is conditionally independent of its non-descendants given its parents. \cite{koller2009probabilistic}
\end{definition}

When dealing with spaces composed solely of discrete-valued random variables, to each node $X_i$ we assign a state $x_i \in Val(X_i)$. We could display the conditional probability distribution $P(X_i | \text{Pa}_{X_i})$ in a conditional probability table (CPT), where
\begin{align*}
\sum_{i \in \{1, \ldots , n\}} P(x_i | \text{Pa}_{X_i}) = 1.
\end{align*}
So, a BN $\G$ exists of a graph with a collection of local probability distributions, displayed in CPTs. Together, these local probability distributions give the joint probability distribution of the BN. When a CPT contains one or more zeros, we deal with determinism. 
\begin{definition}[Deterministic relation]
That is, there exists a function $f: Val(\text{Pa}_{X_i}) \to Val(X_i)$, such that
\begin{align*}
P(x_i | \text{Pa}_{X_i}) =
\begin{cases}
1 & x_i = f(\text{Pa}_{X_i}) \\ 
0 & \text{otherwise}.
\end{cases}
\end{align*} 
\end{definition}
Furthermore, we use $\bfX \subseteq \X$ to denote a set of random variables, while $\bfx$ denotes an assignment of values to the variables in this set. For convenience a state of a BN is denoted as $\bfx = (x_1, \ldots , x_n)$. And $P(\bfx)$ denotes the probability of state $\bfx$.

\vspace{-1pc}
\subsection{Inference}
When values of variables are known, we call this set $\bfE \subset \X$ evidence. We can formulate our main goal as: given a set of evidence $\bfE = \bfe$ and nodes of interest $\bfX \subset \X$ -- such that $\bfE \cap \bfX = \emptyset$ -- what is the probability distribution $P(\bfX | \bfE = \bfe)$? The task of answering this question is called inference. We write $P$ as the posterior probability distribution of interest, with reduced CPTs according to the evidence nodes.

\begin{definition}[Feasible state]
A feasible state of the BN is a state $\bfx$ such that $P(\bfx)>0$, i.e. each unique CPT-value corresponding to state $\bfx$ is positive.
\end{definition}

Since naively summing out all possible configurations -- to determine $P(\bfX | \bfe)$ -- could easily result in an exponential blown up, we want to appeal to a more sophisticated approach. Approximate inference methods provide such a solution. In this article, we focus on MCMC sampling methods.

\vspace{-1pc}
\subsection{MCMC methods}
MCMC sampling methods construct a Markov chain such that, although the first sample may be generated from the prior distributions, successive samples are generated from distributions that provably get closer and closer to the desired posterior distribution. One can examine the desired distribution by observing its equilibrium distribution after a number of steps. In order to use this tempting feature of MCMC methods, we need to guarantee that the limit of this process exists and is unique. Since we only consider Markov chains on finite state spaces, from the theory of Markov chains we know that if a Markov chain is regular and reversible with respect to a distribution $\pi$, then $\pi$ is the unique stationary distribution. These notions are defined below \cite{koller2009probabilistic}.


\begin{definition}[Regular Markov chain]
A Markov chain is said to be regular if there exists some number $h \in \mathbb{N}$ such that, for every $\bfx, \bfx' \in Val(\bfX)$, the probability of getting from $\bfx$ to $\bfx'$ -- denoted as $\bfx \to \bfx'$ -- in exactly $h$ steps, is $> 0$. 
\end{definition}

\begin{definition}[Reversible Markov chain]
A finite-state Markov chain $Q$ is called reversible if there exists a unique distribution $\pi$ such that for all states $\bfx$ and $\bfx'$
\begin{align}
\pi(\bfx) Q(\bfx \to \bfx') = \pi(\bfx') Q(\bfx' \to \bfx).
\end{align}
\end{definition}

\begin{definition}[Stationary distribution]
A distribution $\pi$ is a stationary distribution for a Markov chain Q if
\begin{align}
\pi(\bfx') = \sum_{\bfx \in Val(\bfX)} \pi(\bfx) Q(\bfx \to \bfx').
\end{align}

\end{definition}

Having recalled these definitions from probability theory, we visit the widely used MCMC approximate inference method Gibbs sampling. And -- as already mentioned in the {\normalfont\scshape \sffamily introduction} -- its limitations.

\vspace{-1pc}
\subsection{Gibbs sampling}
Gibbs sampling \cite{geman1984stochastic} is one of the most popular MCMC methods to date. Let $(X_1, \ldots , X_n)$ be an arbitrary ordering of the variables in $\G$. The Gibbs sampling algorithm begins with a random assignment $\bfx^{(0)}$ to all variables in the BN. Then, for $t = 1, \ldots, T$ it performs the following $T \in \mathbb{N}$ steps (each step is called a Gibbs iteration). For $i=1, \ldots, n$, it generates a new value $x_i^{(t)}$ for $X_i$ by sampling a value from the distribution $P(X_i | \bfx_{-i}^{(t)})$, where $\bfx_{-i}^{(t)} = (x_1^{(t)}, \ldots, x_{i-1}^{(t)}, x_{i+1}^{(t-1)}, \ldots, x_{n}^{(t-1)})$. After $T$ samples are generated, all one-variable marginals can be estimated using the following quantity
\begin{align}\label{eq:gibss}
\widetilde{P}_T(x_i) = \frac{1}{T} \sum_{t=1}^T P(x_i | \bfx_{-i}^{(t)} ).
\end{align}
In the limit of infinite samples, $\widetilde{P}_T(x_i)$ will converge to $P(x_i)$ if the underlying Markov chain is regular and reversible. Thus, if the BN contains no deterministic relations, then the Markov chain is guaranteed to have a stationary distribution. Unfortunately, when the BN has deterministic dependencies, regularity and reversibility could break down and the estimation given in Equation (\ref{eq:gibss}) could no longer converge to $P(x_i)$ \cite{venugopal2013giss}. We illustrate this with the following example.
\begin{example}\label{ex:gibbs}
In Figure \ref{gibbs}, consider the initial configuration $(a^{(0)} = 0, b^{(0)} = 0)$ -- shortened $(0,0)$ -- and suppose no evidence is available. 
In the first Gibbs iteration, we resample both unobserved variables, one at a time, in the order $A, B$. So, we first sample $a^{(1)}$ from the distribution $P(A | B = 0)$. According to the CPTs, with probability $1$ this turns out to be $A=0$. Consecutively, we sample $b^{(1)}$ from the distribution $P(B | A = 0 )$. Which always returns $B = 0$. As a consequence the Markov chain created by Gibbs sampling behaves like
\begin{align*}\label{gibbs-trap}
(0,0) \to (0,0) \to (0,0) \to \ldots ,
\end{align*}
yielding that all samples are equal to $(0,0)$. The real distribution for $A$ on the other hand must equal $P(A=0) = P(A=1) = 0.5$.
\end{example}
In the above example, the main problem of applying Gibbs sampling on BNs with deterministic relations becomes visible: the Markov chain could get trapped in a subset of the entire state space. Therefore, the Markov chain generated by this process does not converge to the desired unique stationary distribution.

\vspace{-1pc}
\subsection{MC-SAT}
Many solutions have been proposed in the past to address the above problem. Notable examples are the Sample Search and GiSS algorithm \cite{venugopal2013giss} or the slice sampling method \cite{besag1993spatial, damlen1999gibbs, gilks1996interdisciplinary}. The algorithm MC-SAT \cite{poon2006sound} is a special case of slice sampling and applies the strategy of using auxiliary variables. Poon and Domingos show -- in the more general framework of Markov Logic Networks (MLNs) -- that MC-SAT is a sound MCMC algorithm, meaning it generates a Markov chain which is regular and reversible, even in the presence of deterministic relations.

To use MC-SAT for BN inference, a BN needs to be converted to a weighted weighted satisfiability problem (SAT). This has two drawbacks. In the first place, an explicit translation of a BN to a weighted SAT problem is memory intensive. Secondly, the graphical dependencies of BNs are lost when the BN structure is translated to a collection of weighted clauses. In order not to suffer from these drawbacks, we used concepts of the MC-SAT algorithm to create an MCMC sampling method that preserves the BN representation.

\vspace{-1pc}
\section{Prune sampling}
In this section we define our main contribution: the \ps algorithm. We start introducing the mathematical notation and definition of \textit{Prune Sampling}. Consecutively, we prove theoretically that \ps always generates a regular and reversible Markov chain with respect to the desired distribution. 

\vspace{-1pc}
\subsection{Definition of {prune sampling}}
\indent Before describing the algorithm, we introduce some additional notation. Given a BN structure $\G$ with $i = 1, \ldots , n$ variables and corresponding CPTs. For $ l_i \in | Val(X_i) | \cdot | Val(\text{Pa}_{X_i})|$, let 
\begin{align}
\begin{split}
\C := \{ & k(l_i) : c_{ k(l_i)} \text{ is a CPT-entry of $X_i$} \}
\end{split}
\end{align}
be the collection of all CPT-labels of $\G$. In general, $k$ is an abbreviation of the name assigned to node $X_i$.
\begin{example} The collection of CPT-labels $\C$ for the BN in Figure \ref{gibbs} contains 6 labels: 2 for the CPT of node $A$ -- indexed by $A(1), A(2)$ -- and 4 for the CPT of node $B$ -- indexed by $B(1), B(2), B(3), B(4)$. For completeness, $\C = \{ A(1), A(2), B(1), B(2), B(3), B(4) \}$.
\begin{center}
\begin{figure}
\begin{tikzpicture}[node distance=1.2cm, >=triangle 60]
\node [events] (a) {A};
\node [events,  right=of a] (b) {B};

\draw [->] (a) -- (b);

{\small
\node [below = 0.5cm of a, xshift = -1cm] (ta) {
    \begin{tabular}{|c||c|} \hline
    	$A=0$ & $0.5$ \hspace{0.5pc} $A(1)$\\ \hline
    	$A=1$ & $0.5$ \hspace{0.5pc} $A(2)$ \\ \hline
    \end{tabular}
    };
    
\node [below = 0.5cm of b, xshift = 1.5cm] (tb) {
    \begin{tabular}{|c||c|c|} \hline
	&$A=0$&$ A=1 $  \\ \hline \hline
	$B=0$ & $1$ \hspace{0.5pc} $B(1)$& $0$ \hspace{0.5pc} $B(2)$  \\ \hline
	$B=1$ & $0$ \hspace{0.5pc} $B(3)$& $1$ \hspace{0.5pc} $B(4)$ \\ \hline
\end{tabular}
    };}
    
\draw [dotted] (a) -- (ta);
\draw [dotted] (b) -- (tb);
\end{tikzpicture}
\caption{A BN with a deterministic relation: the state of B is equal to the state of A with probability 1. In this BN we display the corresponding probability at the left side of the CPT entry and the indexation at the right side.}
\label{gibbs}
\end{figure}
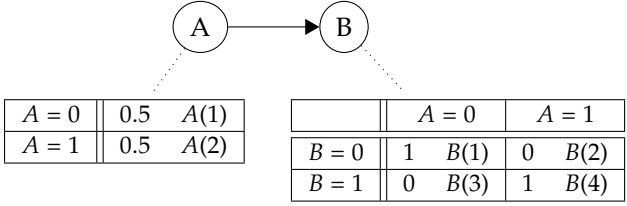
\end{center}
\end{example}
\begin{figure*}[h!]
\centering
\begin{tikzpicture}[node distance=1.2cm, >= triangle 60]

\node [events] (kidney) {Kidney};
\node [events, below right=of kidney] (bp) {BloodPres.};
\node [events, above right=of bp] (lifestyle) {Lifestyle};
\node [events, below=of bp] (measure) {Measurement};
\node [events, below right=of lifestyle] (sports) {Sports};

\draw [->] (kidney) -- (bp);
\draw [->] (lifestyle) -- (bp);
\draw [->] (lifestyle) -- (sports);
\draw [->] (bp) -- (measure);

{\small
\node [left = 1cm of kidney] (t_k) {
    \begin{tabular}{|c||c|} \hline
    	$k_b$ & $\cancel{0.5}$\\ \hline
    	$\mathbf{k_g}$ & $\mathbf{0.5}$ \\ \hline
    \end{tabular}
    };
    
\node [right = 1cm of lifestyle] (t_l) {
\begin{tabular}{|c||c|} \hline
	$\mathbf{l_b}$ & $\mathbf{0.5}$\\ \hline
	$l_g$ & $0.5$ \\ \hline
\end{tabular}
};
    
\node [below = 1cm of t_k] (t_bp) {
    \begin{tabular}{|c||c|c|c|c|} \hline
	&$k_b, l_b$ & $ k_b, l_g $ &$ \mathbf{k_g, l_b}$ & $ k_g, l_g $  \\ \hline \hline
	$b_n$ & $\cancel{0.1}$ & $\cancel{0.2}$ & $\cancel{0.2}$ & $0.9$  \\ \hline
	$\mathbf{b_e}$ & $0.9$ & $0.8$ & $\mathbf{0.8}$ & $\cancel{0.1}$ \\ \hline
\end{tabular}};

\node [right = 0.6cm of sports] (t_s) {
    \begin{tabular}{|c||c|c|} \hline
	&$\mathbf{l_b}$&$ l_g $  \\ \hline \hline
	$s_n$ & $0.8$ & $\cancel{0.2}$  \\ \hline
	$\mathbf{s_y}$ & $\mathbf{0.2}$ & $\cancel{0.8}$ \\ \hline
\end{tabular}};

\node [right = 1cm of measure] (t_m) {
    \begin{tabular}{|c||c|c|} \hline
	&$b_n$&$ \mathbf{b_e} $  \\ \hline \hline
	$m_n$ & $0.9$ & $\cancel{0.1}$  \\ \hline
	$\mathbf{m_e}$ & $\cancel{0.1}$ & $\mathbf{0.9}$ \\ \hline
\end{tabular}}

;}
    
\draw [dotted] (kidney) -- (t_k);
\draw [dotted] (lifestyle) -- (t_l);
\draw [dotted] (bp) -- (t_bp);
\draw [dotted] (sports) -- (t_s);
\draw [dotted] (measure) -- (t_m);
\end{tikzpicture}
\caption{A pruned version of the BloodPressure network around the boldfaced initial state $\bfx = (k_g, l_b, b_e, s_y, m_e)$. Note that the lower the value of the CPT-entry, the higher the probability that the index gets pruned. We see that $S_{\C_\bfx^{\text{np}}}$ contains two feasible states, i.e. $(k_g, l_b, b_e, s_y, m_e)$ and $(k_g, l_b, b_e, s_n, m_e)$.}
\label{pruning}
\end{figure*}
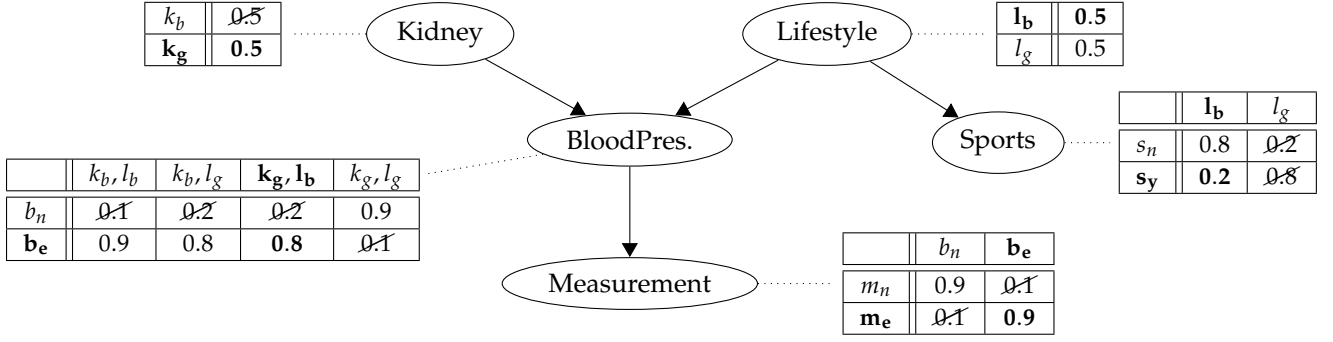
\vspace{-1.5pc}
A state $\bfx$ of the BN corresponds to a unique collection of $n$ CPT-entries. We denote the collection of CPT-labels $k(l_i)$ corresponding to this state by $\C_\bfx$. Accordingly, state $\bfx = (A = 0, B = 0)$ of the BN presented in Figure \ref{gibbs} corresponds to $\C_\bfx = \{ A(1), B(1) \}$. In general, observe that for $i = 1, \ldots, n$
\begin{align*}
P(\bfx) = \prod_{k(l_i) \in \C_\bfx} c_{ k(l_i)}.
\end{align*}

In addition, let $C$ be a collection of CPT-labels. We denote the set of possible (not necessarily feasible) states that correspond to these CPT-labels by $S_{C}$. Having introduced this notation, we are ready to define the concept of \textit{Prune Sampling}.
\begin{definition}[Pruning around state $\bfx$]\label{prunedef}
Let $\C_\bfx^{\text{p}}$ be the subset of $\C$ that is constructed by adding each CPT-label $k(l_i) \in \C \setminus \C_{\bfx}$ with probability $1-c_{k(l_i)}$ to the set $\C_\bfx^{\text{p}}$ and with probability $c_{k(l_i)}$ not. We say that the collection $\C_\bfx^{\text{p}}$ contains the pruned CPT-labels. 
\end{definition}

\begin{example}\label{ex:pruning}
Consider the BN in Figure \ref{pruning}. Pruning around the boldfaced initial state $\bfx = (k_g, l_b, b_e, s_y, m_e)$ could yield the non-crossed indices $\C_\bfx^{\text{p}} = \{ K(2),$ $L(2),$ $BP(4),$ $BP(5), BP(6), S(1), M(1) \}$. Note that the lower the value of the CPT-entry, the higher the probability the label gets pruned. 
\end{example}

The collection of CPT-labels that do not get pruned is given by $\C_\bfx^{\text{np}}:=\C \setminus \C_\bfx^{\text{p}}$. In the situation of Example \ref{ex:pruning}, $S_{\C_\bfx^{\text{np}}}$ exist of two feasible states, i.e. $(k_g, l_b, b_e, s_y, m_e)$ and $(k_g, l_b, b_e, s_n, m_e)$. Having introduced these concepts, one should note three things
\begin{enumerate}
\item $\C_\bfx^{\text{p}}$ is a random set;
\item $\C_\bfx \subset \C_\bfx^{\text{np}} \text{ and } \bfx \in S_{\C_\bfx^{\text{np}}}$;
\item the probability of generating $\C_\bfx^{\text{p}}$ and $\C_\bfx^{\text{np}}$ is given by
\begin{align*}
\prod_{k(l_i) \in \C_\bfx^{\text{p}}}(1-c_{k(l_i)}) \cdot \prod_{k(l_i) \in \C_\bfx^{\text{np}} \setminus \C_{\bfx}} c_{k(l_i)}.
\end{align*}
\end{enumerate}

Due to the pruning of CPT-labels, the number of feasible states in the pruned BN is much smaller in comparison to the number of feasible states in the original BN. This significant decrease of the number of feasible states can make \ps practically applicable. Assuming we have sufficient memory, a breath first search approach can be used to list all feasible states of the pruned BN. From this collection we can easily draw a state uniformly to select the next sample.

\begin{definition}[Uniform sampling over a set of states]
As defined before, $S_{\C_\bfx^{\text{np}}}$ is the set of (feasible) states corresponding to the CPT-labels which are not pruned. We define $\U(S_{\C_\bfx^{\text{np}}})$
as the uniform distribution over the states in $S_{\C_\bfx^{\text{np}}}$ and we write
\begin{align*}
\U(S_{\C_\bfx^{\text{np}}})(\bfy) = \frac{1}{|S_{\C_\bfx^{\text{np}}}|}
\end{align*}
for the probability of sampling state $\bfy$ with respect to this uniform distribution.
\end{definition}
Doing these steps -- pruning, uniform sampling, selecting a new sample -- gives us a sequence of samples that is able to visit the whole state space. We call this process \ps, the pseudo-code could be found in Algorithm \ref{prunealg}. The algorithm takes as input a BN structure $\G$ with corresponding CPTs, an initial configuration $\bfx^{(0)}$ and the integer $T$ for the number of samples to be generated. The algorithm starts with the initial sample $\bfx^{(0)}$. For $t = 1, \ldots , T$ it then prunes around $\bfx^{(t-1)}$ to obtain $\C_{\bfx^{(t-1)}}^{\text{np}}$ and to consecutively sample $\bfx^{(t)}$ from $\U\big( S_{\C_{\bfx^{(t-1)}}^{\text{np}}} \big)$. Finally, the algorithm adds the new sample $\bfx^{(t)}$ to the set $\mathcal{S}$.
\begin{algorithm}
\caption{Prune sampling algorithm}
\label{prunealg}
\begin{algorithmic}
\Function{PruneSampling}{BN, initial, T}
\State $\mathbf{x}^{(0)} \gets $ initial
     \State $\mathcal{S} \gets \{\mathbf{x}^{(0)}\}$
     \For{$t \gets 1 $ to$ $ T}
     \State $\C_{\bfx^{(t-1)}}^{\text{p}} \gets \text{Prune around } \mathbf{x}^{(t-1)}$ \\ \Comment{{\footnotesize See Definition \ref{prunedef}}}
	\State $\C_{\bfx^{(t-1)}}^{\text{np}} \gets \mathcal{C} \setminus \C_\bfx^{\text{p}}$  
     \State $\mathbf{x}^{(t)} \sim  \U(S_{\C_\bfx^{\text{np}}}) $ 
     \State $\mathcal{S} \gets \mathcal{S} \cup \mathbf{x}^{(t)}$
     \EndFor
     \State \Return{$\mathcal{S}$}
\EndFunction
\end{algorithmic}
\end{algorithm}\\
\indent Note that with strictly positive probability we have that $\C_{\bfx^{(t-1)}}^{\text{np}}$ contains all the non-zero indices in $\C$, implying that $S_{\C_{\bfx^{(t-1)}}^{\text{np}}}$ contains all feasible states of the BN. This means that \ps generates a regular Markov chain: with positive probability a state $\bfx$ can transition to an other (arbitrary) feasible state $\bfy$. To show that the Markov chain satisfies the reversibility condition generated by \ps takes more effort and is addressed in the next section.

\vspace{-1pc}
\subsection{A regular and reversible Markov chain}
To make a transition from a state $\bfx$ to a state $\bfy$ we need to prune around $\bfx$ such that non of the labels corresponding to $\bfy$ is pruned. This leads to the following definition. 

\begin{definition}[Pruning around state $\bfx$ and $\bfy$]
Let $\mathcal{C}_{\{\mathbf{x}, \mathbf{y}\}}^{\text{p}}$ be the subset of $\mathcal{C}$ that is constructed by pruning around $\mathbf{x}$ or pruning around $\mathbf{y}$ such that none of the labels corresponding to $\mathbf{x}$ and non of the labels corresponding to $\mathbf{y}$ is contained in $\mathcal{C}_{\{\mathbf{x}, \mathbf{y}\}}^{\text{p}}$.
\end{definition}
The collection of CPT-labels that do not get pruned is given by $\mathcal{C}_{\{\mathbf{x}, \mathbf{y}\}}^{\text{np}} := \C \setminus \mathcal{C}_{\{\mathbf{x}, \mathbf{y}\}}^{\text{p}}$. For each two states $\bfx$ and $\bfy$ there are finitely many ways, $H$, to create a pruned collection $\C_{\{\bfx, \bfy\},h}^{\text{p}}$ and a non-pruned collection $\C_{\{\bfx, \bfy\},h}^{\text{np}}$, where $h =1,\ldots, H$, such that $\bfx$ can make a transition to $\bfy$ by sampling from $\U(S_{\C_{\{\bfx, \bfy\},h}^{\text{np}}})$. We define the transition probability to get from $\bfx$ to $\bfy$ by
{\footnotesize \begin{align}\label{transprob}
{R}_h (\mathbf{x} \to \mathbf{y}) &:= \left(\prod_{k(l_i) \in \C_{\{\bfx, \bfy\},h}^{\text{p}} } (1-c_{k(l_i)}) \right) \cdot \left( \prod_{k(l_i) \in \C_{\{\bfx, \bfy\},h}^{\text{np}} \setminus \mathcal{C}_{\mathbf{x}}  } c_{k(l_i)}   \right) \nonumber \\
& \cdot \mathcal{U} \big( S_{\C_{\{\bfx, \bfy\},h}^{\text{np}}} \big)(\mathbf{y}) \nonumber \nonumber \\ 
&\;=\left(\prod_{k(l_i) \in \C_{\{\bfx, \bfy\},h}^{\text{p}} } (1-c_{k(l_i)}) \right)\cdot \left( \prod_{k(l_i) \in \C_{\{\bfx, \bfy\},h}^{\text{np}} \setminus \mathcal{C}_{\mathbf{x}}  } c_{k(l_i)}   \right)  \nonumber \\
& \cdot \frac{1}{|S_{\C_{\{\bfx, \bfy\},h}^{\text{np}}}|}.
\end{align}}
In words, Equation (\ref{transprob}) expresses the probability of pruning certain CPT-labels around $\bfx$, such that none of the CPT-labels corresponding to $\bfy$ is pruned, and subsequently sampling $\bfy$ uniformly from the states corresponding to the CPT-labels that were not pruned. The total probability of transitioning from $\bfx$ to $\bfy$ is therefore given by
\begin{align} \label{total}
{R}(\mathbf{x} \to \mathbf{y}) = \sum_{h=1}^{H} {R}_h (\mathbf{x} \to \mathbf{y}).
\end{align}
To show reversibility we need to show that the transition probability satisfies the detailed balance equation
$
P(\mathbf{x}) R(\mathbf{x} \to \mathbf{y}) = P(\mathbf{y}) R(\mathbf{y} \to \mathbf{x}) 
$
which equals
\begin{align*}
P(\mathbf{x}) \left(\sum_{h=1}^{H} {R}_h (\mathbf{x} \to \mathbf{y}) \right) = P(\mathbf{y}) \left(\sum_{h=1}^{H} {R}_h (\mathbf{y} \to \mathbf{x}) \right).
\end{align*}
So, it is sufficient to show that
\begin{align} \label{toshow}
P(\mathbf{x}) {R}_h (\mathbf{x} \to \mathbf{y}) = P(\mathbf{y})  {R}_h (\mathbf{y} \to \mathbf{x}), 
\end{align}
for $h = 1, \ldots, H$. The following computation shows that Equation \eqref{toshow} holds
{\footnotesize \begin{align*}
&P(\mathbf{x}) {R}_h (\mathbf{x} \to \mathbf{y}) \\
&= \frac{1}{Z} \cdot P(\mathbf{x}) \cdot \left(\prod_{k(l_i) \in \C_{\{\bfx, \bfy\},h}^{\text{p}} } (1-c_{k(l_i)}) \right)\cdot \left( \prod_{k(l_i) \in \C_{\{\bfx, \bfy\},h}^{\text{np}}  \setminus \mathcal{C}_{\mathbf{x}} } c_{k(l_i)} \right) \\ 
&= \frac{1}{Z} \cdot \prod_{k(l_i) \in \mathcal{C}_{\mathbf{x}}} c_{k(l_i)} \cdot \left(\prod_{k(l_i) \in \C_{\{\bfx, \bfy\},h}^{\text{p}} } (1-c_{k(l_i)}) \right)\cdot \left( \prod_{k(l_i) \in \C_{\{\bfx, \bfy\},h}^{\text{np}} \setminus \mathcal{C}_{\mathbf{x}} } c_{k(l_i)} \right) \\ 
&= \frac{1}{Z} \cdot \left(\prod_{k(l_i) \in \C_{\{\bfx, \bfy\},h}^{\text{p}}} (1-c_{k(l_i)}) \right)\cdot \left( \prod_{k(l_i) \in \C_{\{\bfx, \bfy\},h}^{\text{np}}   } c_{k(l_i)}   \right) \\ 
&= \frac{1}{Z} \cdot \left(\prod_{k(l_i) \in \C_{\{\bfx, \bfy\},h}^{\text{p}}} (1-c_{k(l_i)}) \right)\cdot \left( \prod_{k(l_i) \in \mathcal{C}_{\mathbf{y}}} c_{k(l_i)}   \right) \cdot  \prod_{k(l_i) \in \C_{\{\bfx, \bfy\},h}^{\text{np}} \setminus \mathcal{C}_{\mathbf{y}}  } c_{k(l_i)} \\ 
&= \frac{1}{Z} \cdot \left(\prod_{k(l_i) \in \C_{\{\bfx, \bfy\},h}^{\text{p}} } (1-c_{k(l_i)}) \right)\cdot P(\mathbf{y}) \cdot\left( \prod_{k(l_i) \in \C_{\{\bfx, \bfy\},h}^{\text{np}}  \setminus \mathcal{C}_{\mathbf{y}}  } c_{k(l_i)}   \right) \\ 
&= P(\mathbf{y}) {R}_h (\mathbf{y} \to \mathbf{x}),
\end{align*}}
\noindent where $Z = |S_{\C_{\{\bfx, \bfy\},h}^{\text{np}} }|$. We conclude that \ps generates a regular and a reversible Markov chain with respect to the desired distribution $P$. As discussed before, we now know that $P$ is the unique stationary distribution of the Markov chain generated by \textit{Prune Sampling}.

\vspace{-1pc}
\section{Practical implementation}
To implement the \ps algorithm, two non-trivial steps are required
\begin{enumerate}
\item to generate an initial state of the BN;
\item to sample uniformly over the pruned BN, i.e. sampling from the distribution $\U(S_{\C_\bfx^{\text{np}}})$.
\end{enumerate}
In the next two subsections we elaborate on how we meet these requirements. 

\vspace{-1pc}
\subsection{Generate initial states}
To create an initial state, we need a complete assignment to the BN variables. To fulfill this, MC-SAT uses local search SAT solvers. The first version of the \ps algorithm generates an initial state by the commonly used method {forward sampling}. First, we describe this {forward sampling} method. Subsequently, we present two easy to implement variations.
\begin{definition}[Forward sampling]
Let $X_1, \ldots , X_n$ be a topological ordering of $\X$. We assign a value to the variables consistent with the topological order of the BN. We start assigning randomly values to the root variables of the network. Consecutively, we choose -- conditioned on the assigned values to the variable its parents -- randomly a value from the corresponding distribution defined in the CPT. Continuing this process results in an initial state $\bfx^{(0)}$ of the BN.
\end{definition}
Since forward sampling is guided by its CPT-entries, it is likely that a generated initial state of the BN is a state with high probability. This bias could be a disadvantage of forward sampling. To obtain more diversity in the generated samples, one could consider a custom forward sampling strategy. We propose \textit{random forward sampling}.
\begin{definition}[Random forward sampling]
Suppose we apply forward sampling, but instead of sampling variable $X_i$ from $P(X_i \mid \text{Pa}_{X_i} = \mathbf{x})$, we choose a random sample from the set of non-zero probability states $\{x_i : P(X_i = x_i \mid \text{Pa}_{X_i}=\mathbf{x}) > 0\}$. At this point, $\bfx$ is a collection of $i-1$ assigned values to corresponding variables in the BN. 
\end{definition}
Note that in random forward sampling it is only relevant to know whether a CPT-entry is zero or non-zero. We could also consider a \textit{hybrid forward sampling approach} that combines {forward sampling} and {random forward sampling}.
\begin{definition}[Hybrid forward sampling]
Consider a hybrid approach in which we apply forward sampling, but at each variable $X_i$ either (say with probability $p$) we choose the sampling distribution $P(X_i \mid \text{Pa}_{X_i} = \mathbf{x})$ or (with probability $1-p$) we choose the uniform distribution over $\{x : P(X_i = x \mid \text{Pa}_{X_i}=\mathbf{x}) > 0\}$. 
\end{definition}
Hybrid forward sampling provides a heuristic to assess the maximum a posteriori (MAP) estimate, which is a state $\bfx$ such that $P(\bfx)$ is maximised. In the case one wants to start with a Markov chain from a highly probable state, this approach could be of added value. For situations in which this MAP estimate is useful, could be found in \cite{park2002using}. 


\vspace{-1pc}
\subsection{Sampling from the pruned network}
The MC-SAT algorithm uses an intelligent heuristic based on a weighted SAT problem and applies the concept of simulated annealing \cite{wei2004towards} to generate state nearly uniform (instead of completely uniform). \Ps does generates states completely uniform. This could be considered as a major difference between MC-SAT and prune sampling. In this section we explain two methods how the \ps algorithm samples from $\U(S_{\C_\bfx^{\text{np}}})$. In the first version of \textit{Prune Sampling}, we only implemented the first discussed method.

As argued earlier, where exhaustive listing of all feasible states of the original BN might be impossible due to too much memory consumption, the exhaustive listing of all solutions of the pruned BN is possible. On top of that, drawing a state uniformly from the pruned network even guarantees convergence. Though, the exhaustive enumeration of all feasible states of the pruned network is unavoidable. 


If one still runs into memory problems or one wants to reduce the computational effort, heuristic methods can be developed. We propose to use random forward sampling to construct a set $V$ (of predetermined fixed size) of feasible states of the pruned BN. Subsequently, a state from $V$ can be sampled uniformly. This can be interpreted as a trade-off between uniformity and computational effort. We expect that there should be more intelligent heuristics to obtain near-uniform samples from the pruned network and we again refer to the simulated annealing approach in \cite{wei2004towards} for inspiration.

\begin{figure*}[h!]
\centering
\begin{subfigure}{.49\textwidth}
  \centering
  \includegraphics[width=\linewidth]{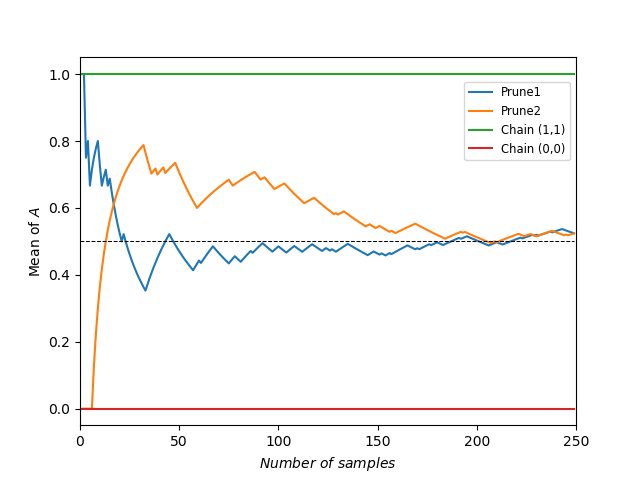}
  \caption{250 samples prune vs Gibbs}
  \label{fig:sub1}
\end{subfigure}
\begin{subfigure}{.49\textwidth}
  \centering
  \includegraphics[width=\linewidth]{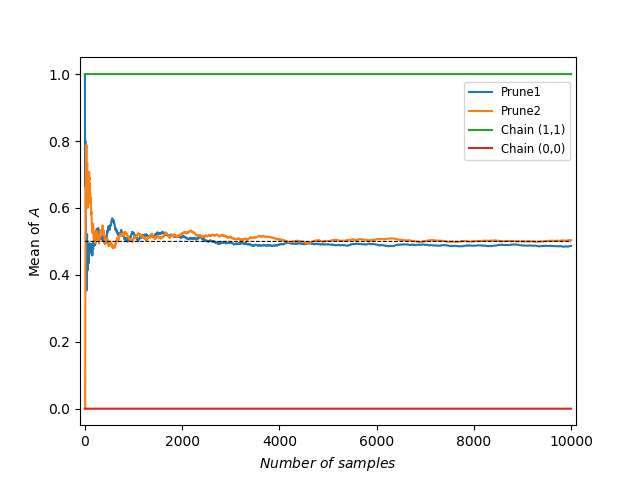}
  \caption{10.000 samples prune vs Gibbs}
  \label{fig:sub2}
\end{subfigure}
\caption{Superior performance of \textit{Prune Sampling}. As described in Example \ref{gibbs}, due to the presence of determinism, Gibbs sampling (green and red line) does not converge to the correct distribution since it is trapped in a subset of the state space.} 
\label{simple-deterministic}
\end{figure*}

\vspace{-1pc}
\section{Experiments}
We compare \ps with the two widely used MCMC inference methods Gibbs- and Metropolis sampling in terms of accuracy, rate of convergence and time consumption. Since \ps is devised to deal with determinism, we conduct experiments on BNs with gradually increasing rates of deterministic relations. On top of that, in order to examine \ps as a broad-applicable BN inference technique, we study the performance of the pruning technique on non-deterministic BNs with increasing rates of available evidence. We used the three MCMC sampling methods to approximate one-variable marginals on small, medium and large BNs from four benchmark domains: simple deterministic-, block shaped-, Grid- and real world BNs. We worked with a tool capable of creating and translating BNs in GeNIe into an equivalent model in Python. In Python, we used the implementations of Gibbs- and Metropolis sampling from the PyMC package. All the BNs used in this study can be found for free online via the bnlearn Bayesian network repository or the UAI repository. We produced results without thinning. If we used a burn-in period, this could be mentioned from the starting number of the 'number of samples' at the x-axis. We executed our experiments on an Intel(R) Core(TM) i5-5300 CPU 2.30GHz core machine with 8 GB RAM, running operating system Windows 10. 

In the section {\normalfont\scshape \sffamily results}, we elaborate on the characteristics of the four benchmark BNs and present the results for accuracy. Details about the average Hellinger distance (a measure for accuracy), the results for the rate of convergence and time consumption, is discussed in the consecutive section {\normalfont\scshape \sffamily performance indicators}. 

\vspace{-1pc}
\subsection{Results}
\begin{figure*}[h!]
\centering
\begin{subfigure}{.49\textwidth}
  \centering
  \includegraphics[width=\linewidth]{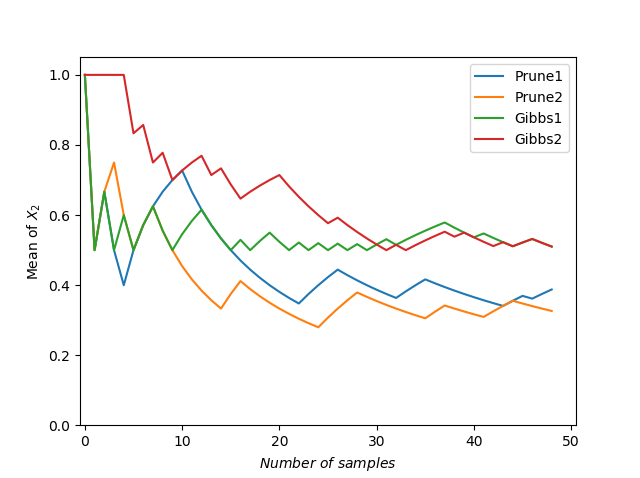}
  \caption{50 samples prune vs Gibbs}
  \label{fig:sub1}
\end{subfigure}
\begin{subfigure}{.49\textwidth}
  \centering
  \includegraphics[width=\linewidth]{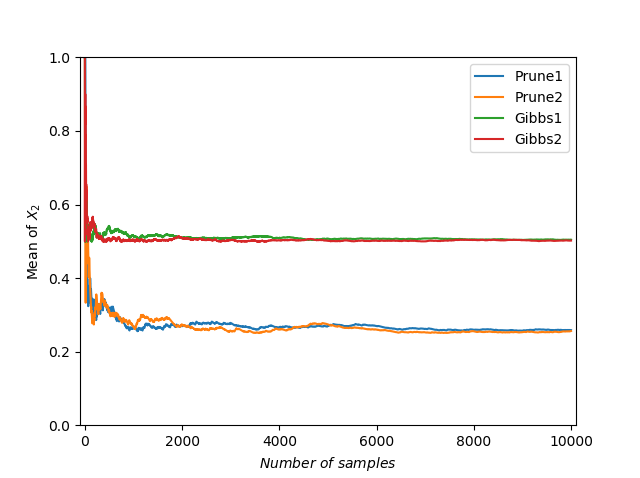}
  \caption{10.000 samples prune vs Gibbs}
  \label{fig:sub2}
\end{subfigure}
\vspace{0.75pc}
\caption{Beside determinism, a non-deterministic block shaped BN can prevent a Markov chain generated by Gibbs sampling of visiting the entire state space. Again, \ps does converge to the desired distribution.} 
\label{block-BN}
\end{figure*}

\begin{figure*}[h!]
\centering
\begin{subfigure}{0.66\columnwidth}
\includegraphics[width=1.1\columnwidth]{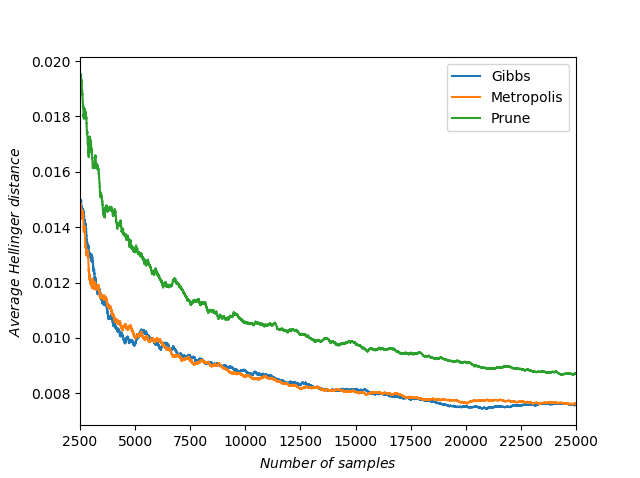}
\caption{Asia 0\% determinism}%
\label{asia}%
\end{subfigure}\hfill%
\begin{subfigure}{0.66\columnwidth}
\includegraphics[width=1.1\columnwidth]{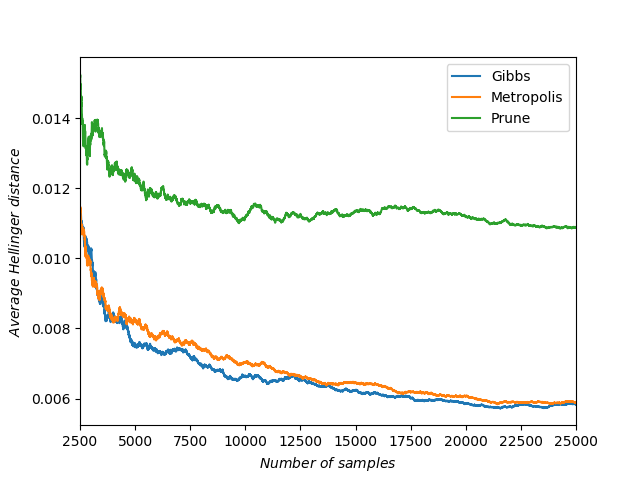}
\caption{Alarm 0\% determinism}%
\label{alarm}%
\end{subfigure}\hfill%
\begin{subfigure}{0.66\columnwidth}
\includegraphics[width=1.1\columnwidth]{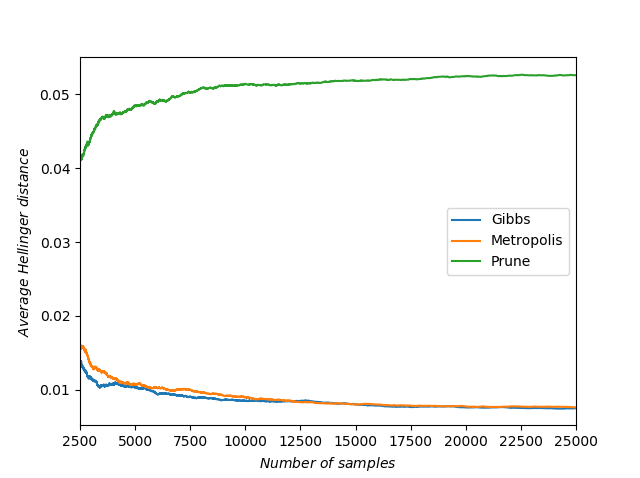}
\caption{Win95pts 0\% determinism}%
\label{win95pts}%
\end{subfigure}\hfill%

\begin{subfigure}{0.66\columnwidth}
\includegraphics[width=1.1\columnwidth]{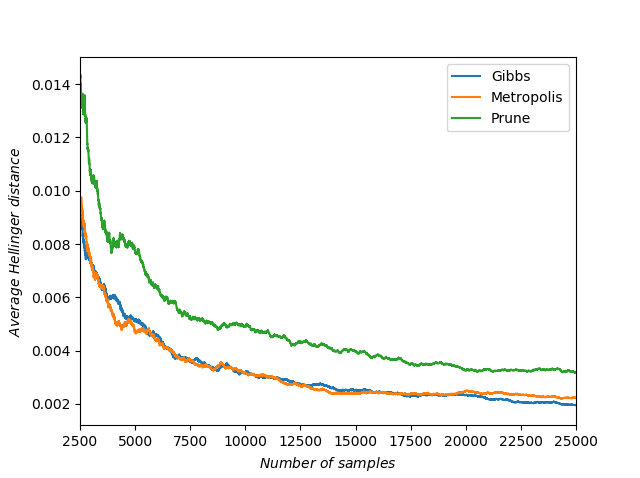}
\caption{Asia 25\% evidence}%
\label{asia_ev}%
\end{subfigure}\hfill%
\begin{subfigure}{0.66\columnwidth}
\includegraphics[width=1.1\columnwidth]{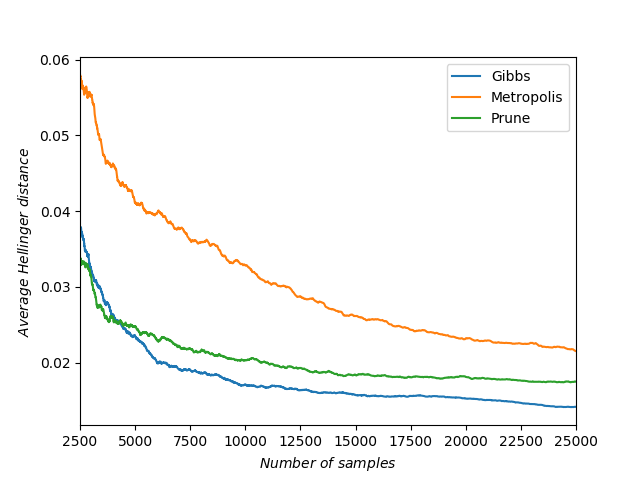}
\caption{Alarm 25\% evidence}%
\label{alarm_ev}%
\end{subfigure}\hfill%
\begin{subfigure}{0.66\columnwidth}
\includegraphics[width=1.1\columnwidth]{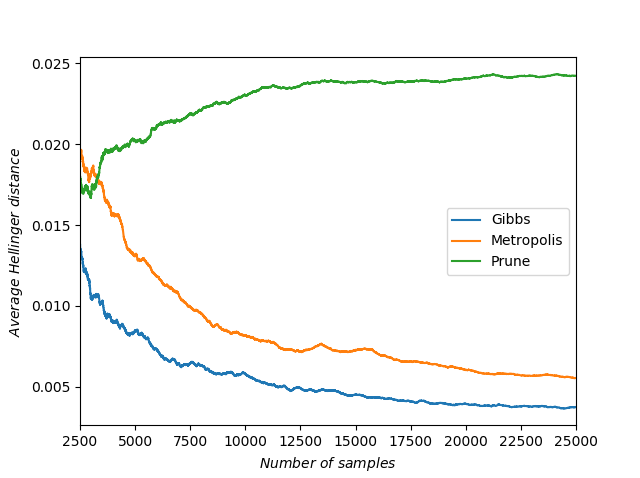}
\caption{Win95pts 25\% evidence}%
\label{win95pts_ev}%
\end{subfigure}\hfill%

\begin{subfigure}{0.66\columnwidth}
\includegraphics[width=1.1\columnwidth]{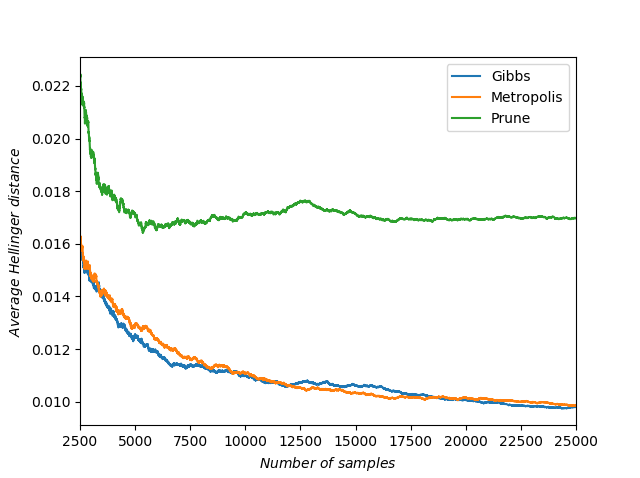}
\caption{Grid 3x3 50\% determinism}%
\label{grid_3x3}%
\end{subfigure}\hfill%
\begin{subfigure}{0.66\columnwidth}
\includegraphics[width=1.1\columnwidth]{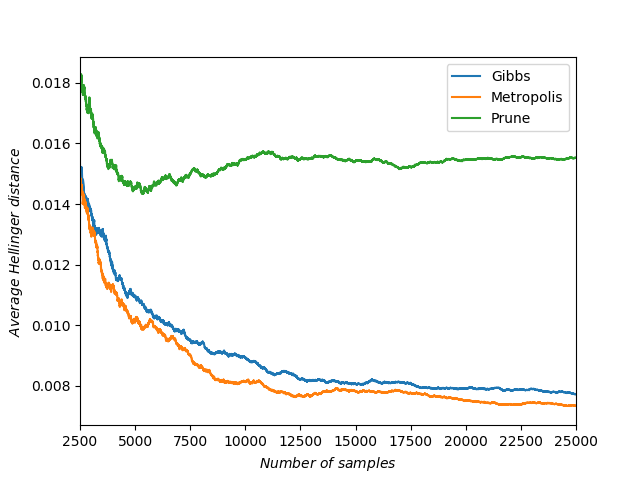}
\caption{Grid 5x5 50\% determinism}%
\label{grid_5x5}%
\end{subfigure}\hfill%
\begin{subfigure}{0.66\columnwidth}
\includegraphics[width=1.1\columnwidth]{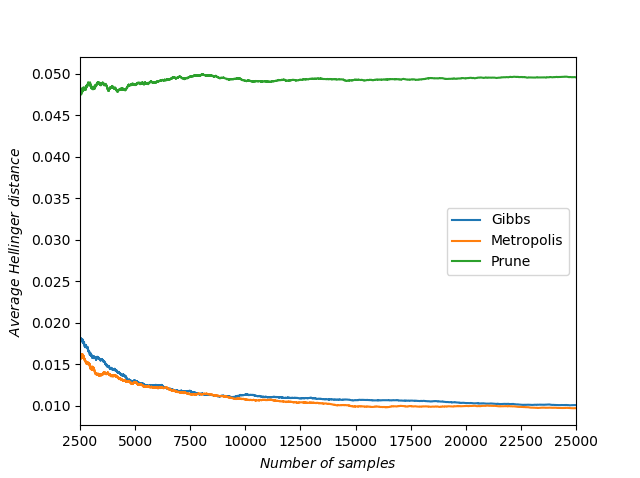}
\caption{Grid 8x8 50\% determinism}%
\label{grid_8x8}%
\end{subfigure}\hfill%

\vspace{0.75pc}
\caption{Average Hellinger distance between the exact and the approximate one-variable marginal plotted as a function of the number of samples generated by Gibbs-, Metropolis- and \textit{Prune Sampling}. This first version of \ps is not competitive with conventional MCMC approximation methods in term of accuracy.}
\label{results1}
\end{figure*}

Figure \ref{simple-deterministic}-\ref{results1} and Table \ref{ROC-table}-\ref{time-table} show the results.  On first sight, we can see that \ps is a less accurate but fast approximation method for small and medium sized BNs. For large BNs, \ps is a less accurate and time intensive method. \\

\textbf{Simple deterministic BN.} As illustrated in Example \ref{ex:gibbs}, we know that a Markov chain generated by Gibbs sampling could be trapped in a subset of the state space and therefore does not always converge to the desired distribution. Based on this deterministic BN (Figure \ref{gibbs}), Figure \ref{simple-deterministic} displays the trapped and converging Markov chains generated by Gibbs- and \textit{Prune Sampling} respectively. This plot shows the ratio that the Markov chain assigns the most common state to the variable of interest. The horizontal red and green lines at 0 and 1 correspond to the trapped Markov chains generated by Gibbs sampling. The red line represents: $P(A=1)=0$. The green line represents: $P(A=1)=1$. However, if we apply \ps on this example, from Equation (\ref{total}) we could derive that
\begin{align*}
Q((0,0) \to (0,0)) = \frac{1}{2} \cdot \frac{1}{2} + 1 \cdot \frac{1}{2} = \frac{3}{4}.
\end{align*}
This means that \ps is capable of making a transition from $(0,0)$ to $(1,1)$ and vice versa. In Figure \ref{simple-deterministic}, the two moving lines -- blue and orange -- correspond to the trace plots of two Markov chains generated by \textit{Prune Sampling}. From both initial states -- $(0,0)$ and $(1,1)$ -- the \ps algorithm is able to move around the entire state space, i.e. able to converge to the correct mean. The horizontal line at $0.5$ represents the exact probability of variable $A=0$. \vspace{0.1pc}

\textbf{Block shaped BN.} Not only determinism could prevent Gibbs sampling of visiting the whole state space. Consider the BN $X_1 \to X_2 \to \ldots \to X_n$, where each $X_i \in \{0,1,2,3\}$ for $1 \leq i \leq n$. Let $X_1$ be uniformly distributed and let each $X_i$, for $i = 2, \ldots , n$ be conditionally distributed according to the block shaped distribution $P(X_i |X_{i-1})=$
\begin{table}[h]\resizebox{\columnwidth}{!}{
  \centering
\begin{tabular}{|c||c|c|c|c|} \hline
	&$X_{i-1}=0$ & $ X_{i-1}=1 $ & $X_{i-1}=2$ & $X_{i-1}=3$  \\ \hline \hline
	$X_i =0$ & $0.5$ & $0.5$ & $0$ & $0$ \\ \hline
	$X_i =1$ & $0.5$ & $0.5$ & $0$ & $0$ \\ \hline
	$X_i =2$ & $0$ & $0$ & $0.5$ & $0.5$ \\ \hline
	$X_i =3$ & $0$ & $0$ & $0.5$ & $0.5$ \\ \hline
\end{tabular}
}.
\caption{A block shaped CPT}
\end{table}

Applying Gibbs sampling will not reveal the correct posterior distribution since if $X_1 \in \{0, 1\}$, then for all $i$: $X_i \in \{0, 1\}$, hence the Markov chain is trapped in the subset $\{0,1\}$ of the state space $\{0,1,2,3\}$. In Figure \ref{block-BN} we see that the Gibbs sampling generates two disconnected regions: the Markov chain is trapped in $\{0, 1\}$ or in $\{2,3\}$. The red and green Markov chain find with probability $0.5$ that $X_i = 0$ or $X_i = 1$ and $X_i = 2$ or $X_i = 3$ respectively. In the same figure we see that the Markov chain generated by \ps -- the blue and orange line -- is again able to move freely around the entire state space and therefore converges -- for all $2 \leq i \leq n$ -- to the correct probability $P(X_i = k)= 0.25$, where $k \in \{0,1,2,3\}$. \\

\textbf{Real world BNs.} We experimented with three real world Bayesian networks. The small network (8 nodes, 18 parameters) Asia \cite{lauritzen1988local}, the medium network (37 nodes, 509 parameters) Alarm \cite{beinlich1989alarm} and the large network (76 nodes, 574 parameters) Win95pts. Figure \ref{results1}(a)-(c) show the results of experiments with 0\% deterministic relations. In Figure \ref{results1}(d)-(f), we conducted experiments with 25\% available evidence. Figure \ref{results1}(a)-(f) reveal that \ps is a reliable BN inference technique for both deterministic as non-deterministic BNs. Though, we see strong underperformance of Prune Sampling on the win95pts network.. \\

\textbf{Grid BNs.}  We conducted experiments with grid BNs of size $3 \times 3$, $5 \times 5$ and $8 \times 8$. This type of BNs is generated by Sang, Beame, and Kautz 2005 \cite{sang2005solving}. We used Grid networks with 50\% deterministic- and 50\% stochastic relations. In Figure \ref{results1}(g)-(i), we see that \ps does not reach accuracy as Gibbs- and Metropolis do. Hence, in contrast to our expectations, we conclude that \ps is not competitive on the class of BNs with the most available deterministic relations.

\vspace{-1pc}
\section{Performance indicators}
In this section, we make clear how we measured the performance of the sampling methods and present our results in terms of the rate of convergence and time consumption.

\vspace{-1pc}
\subsection{Average Hellinger distance} 
We measured accuracy using the average Hellinger distance (AHD) between the exact and the approximate one-variable marginal. The AHD quantifies the closeness of two probability distributions. For our discrete probability distributions $P$ and $Q$ with $n$ binary variables, the AHD is defined as
\begin{align*}
H(P,Q) = \frac{1}{\sqrt{2}}\sqrt{ \sum_{i=1}^n ( \sqrt{p_i} - \sqrt{q_i})^2 }.
\end{align*}
Note that the maximum distance $H(P,Q) = 1$ occurs when for all $i$: $p_i =1$ and $q_i = 0$ or vice versa. For all our experiments, we first computed all $25.000$ Hellinger distances with respect to the exact probability. Note that this exact probability could be computed by an exact inference algorithm, for example by using the decision modelling software in GeNIe. Consecutively, for all $100$ runs we averaged these Hellinger distances at every $t$-th point in the sample, $1 \leq t \leq 25.000$. For all the 9 benchmark BNs, this yielded the AHDs displayed in Figure \ref{results1}.

\vspace{-1pc}
\subsection{Rate of convergence} 
We devised a procedure to obtain the performance of the MCMC sampling methods in the limit of infinite simulation time. In this subsection, we explain how relatively short simulations -- $100$ runs of $25.000$ samples -- could be used to determine the rate of convergence (ROC).

\begin{figure*}[h!]
\centering
\begin{subfigure}[t]{\columnwidth}
  \centering
  \captionsetup{width=.8\columnwidth}
  \includegraphics[width=\columnwidth]{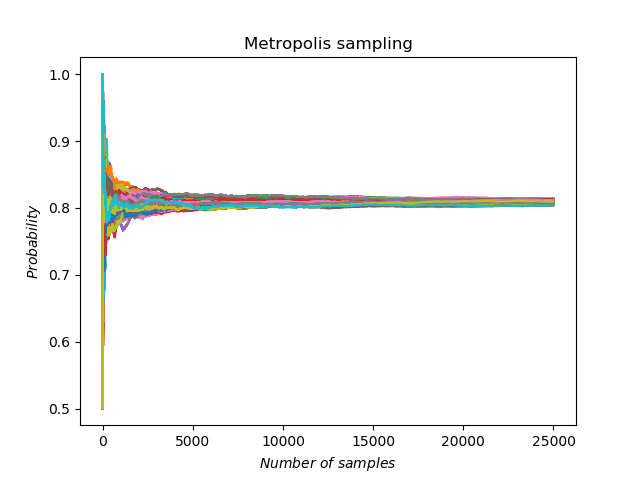}
  \caption{On the Grid $8 \times 8$ BN, a bunch of $100$ Metropolis runs of $25.000$ samples, approximates the one-variable marginal $P(X_{8,8} = T) \approx 0.81$.}
  \label{sub_a}
\end{subfigure}
\begin{subfigure}[t]{\columnwidth}
  \centering
  \captionsetup{width=.8\columnwidth}
  \includegraphics[width=\linewidth]{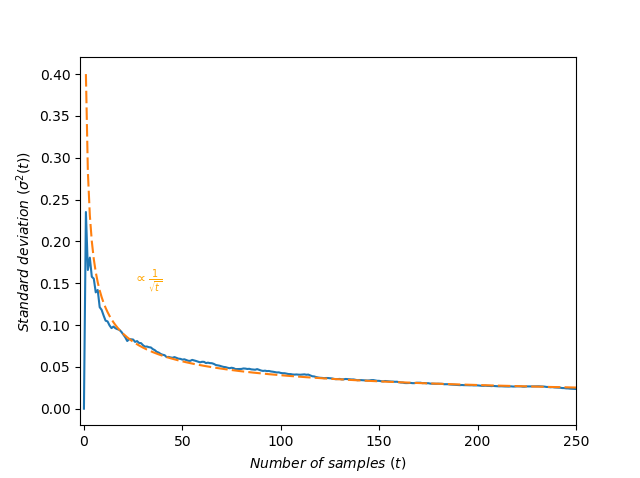}
  \caption{According to convergence class $\mathcal{O}(t^{-1/2})$, the standard deviation ${\sigma}_t^2$ of the $100$ Metropolis runs decreases with the number of samples $t$ with rate $t^{-1/2}$.}
  \label{sub_b}
\end{subfigure}

\begin{subfigure}[t]{\columnwidth}
  \centering
  \captionsetup{width=.8\columnwidth}
  \includegraphics[width=\columnwidth]{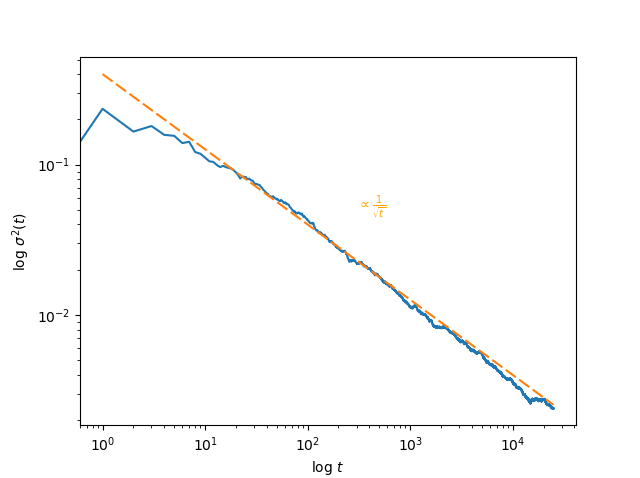}
  \caption{Plotting the $\log$ of the standard deviation -- $\log {\sigma}_t^2$ -- versus the $\log$ of the number of points -- $\log t$ -- yields a linear function, which could be approximated.}
  \label{sub_c}
\end{subfigure}
\begin{subfigure}[t]{\columnwidth}
  \centering
  \captionsetup{width=.8\columnwidth}
  \includegraphics[width=\columnwidth]{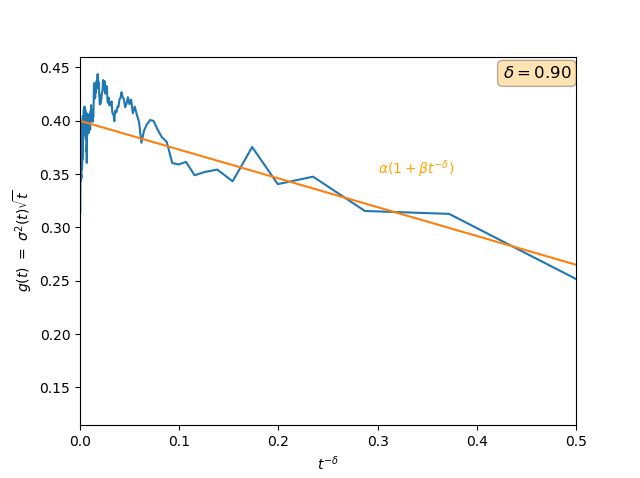}
\caption{To determine $\alpha$ we ignore the interval $10^0$ - $10^1$ in Figure \ref{sub_c}. In doing so, we introduce a polynomial expansion to approximate the linear log plot as $\alpha(1+\beta t^{-\delta})$. If one fits this function to the above line, we find $\alpha \approx 0.40$.}
  \label{sub_d}
\end{subfigure}
\caption{Procedure to determine the proportionality constant of a MCMC method by approximating the asymptotic behavior of the standard deviation.}
\label{determine-c}
\end{figure*}

Say, we qualify the probability we want to know as the expected value of a random variable $Y$, such as $\mu = \E[Y]$. Suppose that by repeating a MCMC sampling method, we generate $N$ runs: $\bfy_1, \ldots , \bfy_N$. Where each run (again) exists of $T$ number of samples. In order to approximate $Y$, we could take the average $\hat{\mu} = \sum_{s=1}^N \bfy_s$. The accuracy of this approximation depends on the number of runs $N$ and the number of samples $T$. A possible measure for the error of MCMC approximation methods, is the standard deviation ${\sigma}_t^2 = \langle y_t^2 \rangle - \langle y_t \rangle ^2$, where
\begin{align*}
\centering
\langle y_t \rangle = \frac{1}{N} \sum_{s=1}^N \bfy_s^{(t)} \hspace{1pc} \text{and} \hspace{1pc}  \langle y_t^2 \rangle = \frac{1}{N} \sum_{s=1}^N (\bfy_s^{(t)})^2,
\end{align*}
and $1 \leq t \leq T$. So, $\bfy_s^{(t)}$ denotes the $t$-th element in the sample during the $s$-th time we run the MCMC method. As an example, in Figure \ref{determine-c}(a) we have run metropolis sampling $N=100$ times to generate samples of $T = 25.000$ points. It could be shown \cite{owen2013monte} that $\sigma_t^2$ decreases with the number of samples $t$ according to
\begin{table}[t]
\begin{center}
\begin{tabular}{l c c c}  
\toprule
\multicolumn{4}{r}{Sampling method} \\
\cmidrule(r){2-4}
Bayesian \\ network    & Gibbs    & Metropolis & Prune  \\
\midrule
Asia\_ev0 & 0.52 & \textbf{0.51} & 0.81  \\
Alarm\_ev0 & \textbf{0.43} & 0.46 & 0.79  \\
Win95pts\_ev0 & \textbf{0.48} & 0.55 & 0.59  \\
Asia\_ev25 & 0.47 & \textbf{0.45} & 0.77  \\
Alarm\_ev25 & 0.40 & \textbf{0.37} & 0.49  \\
Win95pts\_ev25 & \textbf{0.45} & 0.48 & 0.50  \\
Grid 3x3 & \textbf{0.37} & 0.38 & 0.62  \\
Grid 5x5 & 0.54 & \textbf{0.49} & 0.55  \\
Grid 8x8 & \textbf{0.39} & 0.40 & 0.53  \\
\bottomrule
\end{tabular}
\caption{Proportionality constants of the ROC for Gibbs-, Metropolis- and \ps based on $100$ runs of $25.000$ samples.}
\label{ROC-table}
\end{center}
\end{table}
\begin{flalign*}
{\sigma}_t^2 \propto \frac{1}{\sqrt{t}}. 
\end{flalign*}
To emphasize that the rate of convergence is of order $t^{-1/2}$ and to de-emphasize $\sigma_t$, we write ROC = $\mathcal{O}(t^{-1/2})$ as $t \to \infty$. In Figure \ref{determine-c}(b), we see that -- based on $N = 100$ simulations -- $\sigma_t$ indeed behaves like $t^{-1/2}$.
One could consider $\sigma_t^2 = \alpha / \sqrt{t}$. This proportionality constant $\alpha$ quantifies the ROC of the MCMC sampling technique that belongs to the convergence class  $\mathcal{O}(t^{-1/2})$. Therefore, $\alpha$ is a performance indicator of the MCMC sampling techniques we are examining in contrast.

In order to determine this constant $\alpha$ we introduce an intelligent procedure. First, plot $\log \sigma_t$ versus $\log t$. In doing so, we obtain Figure \ref{determine-c}(c). For the interval $10^0 < t < 10^1$, we see that $\sigma_t$ does not yet behave as $t^{-1/2}$. Ideally -- when determining $\alpha$ for the simulations in this figure -- we do not involve this non-representative region. We show how to ignore this region in a sophisticated way. First, we introduce an auxiliary polynomial expansion such that
\begin{align*}
{\sigma}_t^2 = \frac{\alpha}{\sqrt{t}}(1+\beta_1 t^{-\delta} + \beta_2 t^{-2\delta} + \ldots )\ .
\end{align*}
Due to the prospect of overfitting, we simplify the above equation as
\begin{align*}
g(t) = {\sigma}_t^2 \sqrt{t} = \alpha(1+ \beta t^{-\delta})\ .
\end{align*}
One should note that if we plot $g(t)$ in terms of $t^{-\delta}$, for certain $\delta$, $g(t)$ becomes approximately a linear function. We learn how to choose $\delta$ by doing. Once, we have found an approximate linear plot - like Figure \ref{determine-c}(d) -- where we have chosen $\delta = 0.90$ -- we could interpret $\alpha$ as the point of intersection with the y-axis. Hence, for the bunch of Metropolis samples displayed in Figure \ref{determine-c}(a) we determine that the proportionality constant of the ROC is $\alpha = 0.40$.

Following this procedure, we have determined the ROCs of Gibbs-, Metropolis- and \ps for all the 9 benchmark networks in Figure \ref{results1}. The results are presented in Table \ref{ROC-table}. We see that \ps has always a significantly higher ROC than Gibbs- and Metropolis sampling . Hence, \ps convergences slower to the desired distribution.


\vspace{-1pc}
\subsection{Time consumption}
The last performance indicator we take into account is the time consumption of the MCMC sampling methods. Based on the average of $100$ simulations, for all networks in Figure \ref{results1} we examined the time the methods took to achieve $\sigma_t^2 = 0.01$. Note that the equilibrium a sampling method is converging to, is not necessarily the correct probability (distribution). And, notice that with the information about $\alpha$ -- from Table \ref{ROC-table} -- we can determine the amount of samples a method need to generate to achieve $\sigma_t^2 = 0.01$, namely: $t = (\alpha / 0.01)^{2}$. Consecutively, we measured the time the methods needed to generate this number of samples, including the generation of an initial state at the start. The results are displayed in Table \ref{time-table}. For small and medium sized BNs, we see that \ps is always faster than Gibbs- and Metropolis sampling. For large networks, this is not the case. 
\begin{center}
\begin{table}[!htb]
\begin{center}
\begin{tabular}{l c c c}  
\toprule
\multicolumn{4}{r}{Sampling method} \\
\cmidrule(r){2-4}
Bayesian \\ network    & Gibbs    & Metropolis & Prune  \\
\midrule
Asia\_ev0 & 2.72 & 1.31 & \textbf{0.53}  \\
Alarm\_ev0 & 12.27 & 3.94 & \textbf{3.56}  \\
Win95pts\_ev0 & 36.38 & \textbf{21.32} & 49.85  \\
Asia\_ev25 & 2.56 & 1.05 & \textbf{0.41}  \\
Alarm\_ev25 & 9.92 & 3.07 & \textbf{2.83}  \\
Win95pts\_ev25 & 29.65 & \textbf{16.03} & 40.03  \\
Grid 3x3 & 1.67 & 1.76 & \textbf{0.73}  \\
Grid 5x5 & 12.13 & 4.70 & \textbf{2.42}  \\
Grid 8x8 & 20.50 & \textbf{8.83} & 105.17  \\
\bottomrule
\end{tabular}
\caption{Time consumption of Gibbs-, Metropolis- and \ps in seconds to achieve $\sigma_t^2 = 0.01$. Based on the standard deviation of $100$ samples.}
\label{time-table}
\end{center}
\end{table}
\end{center}
Due to the exhaustive enumeration during the uniform sampling step, for large BNs this could still result in an exponential blow up of $|S_{\C_\bfx^{\text{np}}}|$. As a consequence, for large BNs \ps becomes a time intensive technique.
\section{Conclusion}
\Ps is a broad applicable approximate MCMC sampling method for all types of BNs which always converges to the desired distribution. For this reason, \ps outperforms Gibbs sampling for a class of block shaped and deterministic BNs. However, this tempting feature comes at a price. If Gibbs- and Metropolis sampling do converge to the correct distribution and regardless of the amount of available evidence or deterministic relations, \ps is a less accurate method with a lower ROC. Though, on small and medium sized BNs \ps is the fastest method. For large BNs, due to exhaustive enumeration of all possible feasible states, \ps becomes a time intensive method. In order to improve this first version of \textit{Prune Sampling}, we advise to develop an intelligent heuristic which avoids a breath first search approach but that does guarantee uniformly sampling from the entire sample space.

\vspace{-1pc}
\subsection{Acknowledgments}
We explicitly thank Gerard Barkema from the department of Computer science, Utrecht University to provide us with the procedure to obtain the performance of the MCMC sampling methods in the limit of infinite simulation time, extrapolated from relatively short simulations. Beside that, this work was supported by the Dutch Ministry of Defence (grant number V1408). 


\end{document}